\documentstyle[11pt]{article}

\newcommand{\bee}{\begin{equation}}
\newcommand{\eee}{\end{equation}}
\newcommand{\bea}{\begin{eqnarray}}
\newcommand{\eea}{\end{eqnarray}}
\newcommand{\hL}{\hat{L}}
\newcommand{\hPhi}{\hat{\Phi}}
\newcommand{\bN}{\bar{N}}
\newcommand{\mubine}{$\mbox{\boldmath$\mu$}$}
\newcommand{\tl}{\tilde{l}}
\begin{document}
\begin{titlepage}

\begin{flushright}
CERN-TH/97-333\\
LPTHE-ORSAY 97/xx  \\
Saclay T97/xx \\
hep-ph/9711517 \\
\end{flushright}

\vskip.5cm
\begin{center}
{\huge{\bf Quark flavour conserving violations of the lepton number }}
\end{center}
\vskip1.5cm

\centerline{ P. Bin\'etruy $^{a,b}$, E. Dudas  $^{a,b}$, S. Lavignac $^{a,c}$
\footnote{ATER at Universit\'e Paris 7}
and C.A. Savoy $^{d,b}$}
 
\vskip 15pt
\centerline{$^{a}$ Laboratoire de Physique Th\'eorique et Hautes Energies
\footnote{Laboratoire associ\'e au CNRS-URA-D0063.}}
\centerline{B\^at. 210, Univ. Paris-Sud, F-91405 Orsay Cedex, {\sc France}}
\centerline{$^{b}$ CERN-TH, CH-1211 Geneva 23, {\sc Switzerland}}
\vskip 3pt
\centerline{$^{c}$ Institute for Fundamental Theory, Dept. of Physics}
\centerline{Univ. of Florida, Gainesville FL 32611, {\sc USA}}
\centerline{$^{d}$ CEA-SACLAY, Service de Physique Th\'eorique,   }
\centerline{F-91191 Gif-sur-Yvette Cedex {\sc France}}
\vglue .5truecm

\begin{abstract}
We study supersymmetric models of lepton and baryon number violation
based on an abelian family gauge group. Due to possible lepton-Higgs mixing,
the lepton violating couplings are  related to the
Yukawa couplings and may be generated by them even if they were absent 
in the original theory. Such terms may be dominant and are not given by the 
naive family charge counting rules. This enhancement mechanism can provide 
an alignment between
lepton-number violating terms and Yukawa couplings: as a result they conserve 
quark flavour. 
A natural way of suppressing baryon number violation in this class of
models is also proposed.
\end{abstract}  

\vfill
\begin{flushleft}
November 1997 \\
\end{flushleft}

\end{titlepage} 
 Most  of the phenomenological  discussions on  R-parity violations in
the supersymmetric extensions of the Standard  Model assume that there
is  a  single R-parity  violating  coupling,    at  least to   leading
order.  This is often presented  in parallel  with the situation among
the   Yukawa  couplings where   the top   Yukawa  coupling is  clearly
leading. But  if this argument   has any truth  in it,  any theory  of
fermion   masses should account   for   the relative size of  R-parity
violating couplings    as well.  In   this paper,  we will  assume the
existence of an abelian family   symmetry which explains the  observed
hierarchies and  mixings in the quark  and charged lepton  sectors and
discuss  its    consequences  for the   R-parity  violating  couplings
\cite{BN,BaGNN,BLR,BoGNN,NP}. 

It   is  well-known  that  the simultaneous    presence among R-parity
violating couplings of  unsuppressed couplings violating lepton number
as  well as baryon number leads  to  dangerously fast proton decay. We
will in most  of  what follows assume the    existence of a   discrete
symmetry  such  as  a  baryonic  parity  which ensures   baryon number
conservation: the only  allowed couplings  violate  lepton number.  We
will however relax in  the end  this assumption  and show  that family
symmetries  may   yield  a natural   suppression mechanism   for  such
couplings. 

In the general approach using  an abelian family symmetry to constrain
the order of magnitude  of  Yukawa couplings \cite{FN}, one  describes
the breaking of  the family symmetry  by the small parameter $\epsilon
\equiv <\theta> / M_F$   where $\theta$ is  a  field of  family charge
normalized   to $-1$ and $M_F$ a   typical flavor symmetry scale.  For
example  denoting  by  $\phi_i$ the family  charge   of the superfield
$\hPhi^i$, the  coupling $\hPhi^i \hPhi^j  \hPhi^k$  is not allowed by
the family symmetry if $x_{ijk} \equiv \phi_i + \phi_j + \phi_k \not =
0$, but $\hPhi^i \hPhi^j \hPhi^k  \theta^{x_{ijk}}$ is. Thus, once the
family  symmetry is spontaneously broken  by $<\theta>  \not = 0$, the
superpotential may include 
\begin{equation}
W \ni \lambda_{ijk} \hPhi^i \hPhi^j \hPhi^k,
\end{equation}
with 
\begin{equation}
\lambda_{ijk} \sim \epsilon^{\phi_i + \phi_j + \phi_k}. \label{naive}
\end{equation}

This sort of  naive power counting is actually  not exact if, for some
reason, the low energy fields, which we will denote by $\Phi_i$ do not
coincide with   the fields $\hat   \Phi^i$ of definite  family charges
$\phi_i$. There is a possible  enhancement of the low energy couplings
with  respect to the naive  estimate (\ref{naive}). A standard example
occurs  precisely in the case  of R-parity breaking: the weak doublets
of hypercharge $-1$ and given family charge  may not coincide with the
Higgs  doublet  $H_d$ and  the lepton doublets  $L_i$  of the Standard
Model.  If they do  not correspond  exactly,  the rotation to  the low
energy states --which are not eigenstates of the family symmetry-- may
yield a  different  order of magnitude  for  the low energy  couplings
(\ref{naive})  in      the case  where    the  field   redefinition is
nonholomorphic.    We will    illustrate this  enhancement   mechanism
precisely  on the R-parity   violating couplings and  show  that it is
accompanied with an  alignment of these  couplings along the direction
of diagonal Yukawa couplings. 

\vskip .5cm To be more precise,  let us first write the superpotential
in terms of the  fields which have  definite charges under the abelian
family symmetry: besides   the  standard low energy quark   and lepton
superfields $Q^k$, $U^k$, $D^k$ and $E^k$, the Higgs doublet $H_u$ and
the  Standard  Model singlets  $\bN^k$   of respective  family charges
$q_k,u_k, d_k, e_k,   h_u,  n_k$   ($k=1,2,3$),  we introduce     four
left-handed   doublets $\hL^\alpha$   of hypercharge $-1$   and family
charge $l_\alpha$ ($\alpha=1,2,3,4$). 

The superpotential reads:
\begin{eqnarray}
W &=& \mu_\alpha \hL^\alpha H_u 
+ M_{ij} \bN^i \bN^j \nonumber \\
& & + \lambda^E_{\alpha \beta k} \hL^\alpha \hL^\beta E^k
+ \lambda^D_{\alpha j k} \hL^\alpha Q^j D^k
+ \lambda^N_{\alpha k} \hL^\alpha H_u \bN^k
+ \lambda^U_{i k} Q^i H_u U^k \nonumber \\
& & + \lambda''_{ijk} U^i D^j D^k \label{W1}
\end{eqnarray}
with $\lambda^E_{\alpha \beta k} = - \lambda^E_{\beta \alpha k}$
and $\lambda''_{ijk} = - \lambda''_{ikj}$.
\vskip .3cm

The standard Higgs superfield $H_d$ of hypercharge $-1$ is defined as 
the combination of $\hL^\alpha$ along which $U(1)_Y$ is broken. 

More precisely, introducing $v^\alpha \equiv <~\hL^{\alpha 0}~>$,
\begin{equation}
H_d \equiv {1 \over v_d} \sum_\alpha v_\alpha \hL^\alpha  \label{Hd}
\end{equation}
where $v_d \equiv (\sum_\alpha v_\alpha v^\alpha)^{1/2}$.

The other interesting direction is along \mubine\ $= [\mu_\alpha]$ (if
it is  not completely  aligned with  ${\bf   v} = [v_\alpha]$,  as  is
generally the case). Defining therefore  ${\bf w} = [w_\alpha]$ as the
normed vector  orthogonal  to ${\bf  v}$ in  the ({\bf  v}, \mubine\ )
plane, one introduces the lepton superfield: 
\begin{equation}
L^3 \equiv \sum_\alpha w_\alpha \hL^\alpha.  \label{L3}
\end{equation}
As we will see the corresponding neutrino acquires a mass through the 
$\hL^\alpha H_u$ terms in (\ref{W1}).

We can then write 
\begin{eqnarray}
\hL^\alpha &=& {H_d \over v_d} v^\alpha + L^3 w^\alpha + L^\alpha_\perp 
\nonumber \\
&=& {H_d \over v_d} v^\alpha + {e^\alpha}_i L^i. \label{Lhat}
\end{eqnarray}
where $L^\alpha_\perp = \sum_{a=1,2} 
{e^\alpha}_a L^a$ is orthogonal to ${\bf v}$ and ${\bf w}$ and 
${e^\alpha}_3 = w^\alpha$.
\vskip .3cm

We also introduce the angle $\xi$ between ${\bf w}$ and ${\bf v}$, {\em 
i.e.}
\begin{equation}
\mu_\alpha = \mu \cos \xi {v_\alpha \over v_d} + \mu \sin \xi w_\alpha,  
\label{xi}
\end{equation}

where $\mu \equiv (\sum_\alpha \mu_\alpha \mu^\alpha)^{1/2}$.

The superpotential in (\ref{W1}) now reads:
\begin{eqnarray}
W &=& \mu \cos \xi H_d H_u + \mu \sin \xi L^3 H_u 
+ M_{ij} \bN^i \bN^j \nonumber \\
& & + \lambda^E_{i k} L^i H_d E^k
+ \lambda^D_{i k}  Q^i H_d D^k
+ \lambda^N_{i k} L^i H_u \bN^k
+ \lambda^U_{i k} Q^i H_u U^k \nonumber \\
& & + \lambda_{i j k} L^i L^j E^k
+ \lambda'_{i j k} L^i Q^j D^k
+ \lambda^N_{k} H_d H_u \bN^k
 + \lambda''_{ijk} U^i D^j D^k, \label{W2}
\end{eqnarray}
where
\begin{eqnarray}
\lambda^E_{ik}  \equiv  2 \lambda^E_{\alpha \beta k} {e^\alpha}_i 
{v^\beta \over v_d} \ , \
\lambda^D_{ik}  \equiv  - \lambda^D_{\alpha i k} {v^\alpha \over v_d},  
\nonumber \\
\lambda^N_{k}  \equiv  \lambda^N_{\alpha k}  {v^\alpha \over v_d} \ , \
\lambda_{ijk}  \equiv  \lambda^E_{\alpha \beta k} {e^\alpha}_i 
{e^\beta}_j \ , \nonumber \\
\lambda'_{ijk}  \equiv  \lambda^D_{\alpha jk} {e^\alpha}_i \ , \
\lambda^N_{ik}  \equiv  \lambda^N_{\alpha k} {e^\alpha}_i \ .   
\label{lambda}
\end{eqnarray}
An obvious remark at this point, which will prove to be useful in what 
follows,
is that the baryon number violating couplings $\lambda''_{ijk}$ are
not touched by the field redefinition and therefore stay independent of the 
Yukawa couplings. For the time being, we will set them to zero by imposing 
for example a baryonic parity \footnote{A simple way to obtain a baryonic
parity is by the spontaneous breaking $U(1) \rightarrow Z_N$, which arises
if the field $\theta$ which breaks $U(1)$ has a charge $N$ normalized to the
smallest charge of the theory \cite{IRdiscret}.} .
\vskip .3cm

In the new basis $(H_d, L^i)$, the two neutrinos corresponding to 
$L^a, \; a=1,2$ decouple from the other states and we are left with a 
five-by-five neutralino-neutrino mass matrix  which reads in the 
$(\tilde \gamma, \tilde Z, \tilde H^0_u, \tilde H^0_d, L^0_3)$ basis:
\begin{equation}
M_\nu = \left(
\begin{array}{ccccc}
M_1 c^2_w + M_2 s^2_w & (M_2-M_1) s_w c_w & 0 & 0 & 0 \\
(M_2-M_1) s_w c_w & M_1 s^2_w + M_2 c_w^2 &  m_Z \sin \beta
& -m_Z \cos \beta & 0 \\
0 & m_Z \sin \beta & 0 & -\mu \cos \xi & - \mu \sin \xi \\
0 & -m_Z \cos \beta & -\mu \cos \xi & 0 & 0 \\
0 & 0 & -\mu \sin \xi & 0 & 0
\end{array} \right)
\label{massmat}
\end{equation}
where $c_w \equiv \cos \theta_w$, $s_w = \sin \theta_w$ and $\tan 
\beta \equiv v_u/v_d$.

The non-zero eigenvalue corresponding to the neutrino reads \cite{hempfling}

\begin{equation}
m_{\nu_{3}} = m_0 \tan^2 \xi, \label{eq:mLHu}
\end{equation}
with
\begin{equation}
m_0 = {m_Z^2 \cos^2 \beta (M_1 c_w^2 + M_2 s_w^2) \mu \cos \xi
\over M_1M_2 \mu \cos \xi - m_Z^2 \sin {2 \beta} ( M_1 c_w^2 + M_2 s_w^2)},
 \label{m0}
\end{equation}
where $M_1$ and $M_2$ are the usual $U(1)_Y$ and $SU(2)_L$ gaugino soft 
masses.

As is  well-known \cite{HS,all},  such a   neutrino  mass  is 
compatible   with  the
experimental limits only if the angle $\xi$  is small, that is in case
of approximate  alignment  between ${\bf  v}$  and \mubine  .  We will
consider this situation in what follows. 

The  family  symmetry gives the  order  of magnitude  of the couplings
(\ref{W1}), in the basis of family  symmetry eigenstates. If we assume
that   the  quadratic   terms    are not present   in    the  original
superpotential and  are produced from  the  K\"ahler potential through
the Giudice-Masiero mechanism \cite{GM}, then 
\begin{equation}
\mu_\alpha \sim \tilde m \epsilon^{\tl_\alpha}
\end{equation}
where 
\begin{equation}
\tl_\alpha \equiv |\l_\alpha + h_u|,
\end{equation} 
$\tilde m$  is  a  typical  supersymmetry   breaking mass   scale  and
$\epsilon$ measures  the breaking of the family  symmetry (as usual we
take it to be the sine of the Cabibbo angle).  Let us denote by $\tl_0
\equiv h_d$  the smallest of the $\tl_\alpha$  and assume that  $0 \le
\tl_0 < \tl_i$ ($i=1,2,3$). Then obviously 
\begin{equation}
\mu_i / \mu_0 \sim \epsilon^{\tl_i-\tl_0}.
\label{mudir}
\end{equation}

The components of {\bf v} depend on \mubine\  as well as on the  soft
terms in the scalar potential,
\begin{equation}
V_{{\rm soft}} = (B\mu)_{\alpha} {\hat L}^{\alpha} H_u + m^2_H {H_u^+} H_u
+ m^2_{\alpha \beta}{\hat L}^{\alpha +} {\hat L}^{\beta} + \cdots 
\label{Vsoft}
\end{equation}
where the  order  of magnitude   of   the parameters  are as   follows
\footnote{A  more detailed determination   of the  soft parameters  in
theories with abelian  family symmetries and Giudice-Masiero mechanism
is given in \cite{GDPS}.}, 
\begin{equation}
(B\mu)_\alpha \sim \epsilon^{\tl_\alpha} \ \ \ \ \ \
m^2_H \sim {\tilde{m}}^2 \ \ \ \ \ \
m^2_{\alpha \beta} \sim \epsilon^{|l_\alpha - l_\beta |} \tilde{m}^2.
\label{soft}
\end{equation}
Since $\tl_0 < \tl_i$,  the scalar potential can be minimized  in some
obvious approximations and the $\epsilon$ dependence of  the {\bf v}
components can be easily obtained,   
\begin{equation}
v_0/v_u \sim \epsilon^{\tl_0}\ \ \ \ \ \ \ \ \
v_i/v_0 \sim \epsilon^{\tl_i-\tl_0}.
\label{vdir}
\end{equation}
Thus, within our assumptions,
\begin{eqnarray}
v_d \sim v^0 & \mu \sim \mu_0 \nonumber \\
\hL^0 \sim H_d & \hL^i \sim L_i
\end{eqnarray}
and  $\tan{\beta} = (v_u / v_d ) \sim \epsilon^{- \tl_0}.$ 

{}From (\ref{mudir}) and (\ref{vdir}), the alignement between  {\bf v} and
\mubine\  is controled by the powers $(\tl_i-\tl_0)$, so that
\begin{equation}
{e^\alpha}_i \sim \epsilon^{|\tl_\alpha-\tl_i|}.
\label{eai}
\end{equation}
Since $\sin^2 \xi$ can be written as
\begin{equation}
\sin^2 \xi = {\sum_{\alpha,\beta} (\mu_\alpha v_\beta - 
\mu_\beta v_\alpha)^2 \over2 \mu^2 v^2_d}
\end{equation}
one easily obtains that 
\begin{equation}
\sin^2 \xi \sim \epsilon^{2(\tl_3-\tl_0)}, \label{angle}
\end{equation}
where $\tl_3$ is defined by $0 \le \tl_0 < \tl_3 \le \tl_a$ ($a=1,2$).
One thus checks that the vector 
$w_\alpha$ defined in (\ref{xi}) is of order 
$\epsilon^{|\tl_\alpha - \tl_3|}$ in agreement with $w^\alpha =
{e^\alpha}_3$.

Let us  consider  the consequences of   this mixing for  the  R-parity
violating interactions. The  superpotential (\ref{W1})  is defined  in
terms  of the family  symmetry eigenstates. Its  invariance under this
symmetry  implies that the   couplings are proportional  to powers  of
$\epsilon$  as  given  by (\ref{naive}) if    charge $\phi_i +  \phi_j
+\phi_k \ge 0$, while its analiticity implies that  they vanish if the
total charge  is negative. From this and  the $\epsilon$ dependence of
the mixings in (\ref{eai}), one derives relations between the R-parity
violating  couplings and   the   Yukawa   couplings  as  defined    by
(\ref{lambda}). In the case  where $l_i <   l_0$, the former  come out
larger than the latter in contradiction with experimental limits. It is
possible to escape this conclusion  by assuming sufficiently  negative
$l_i$      charges, so  that    the   couplings   $\lambda^{E}_{ijk}$,
$\lambda^{D}_{ijk}$,   $\lambda^{N}_{ik}$  in (\ref{W1})   vanish   by
analyticity\footnote{If the gauge singlets  $N_i$ are included in  the
theory, this assumption  is  needed in  order to avoid  large neutrino
masses from the seesaw mechanism (cf.  below).}. Then, in all cases, 
\begin{eqnarray}
\lambda_{ijk} & \sim & \epsilon^{\tl_i-\tl_0}\lambda^E_{j k} \ ,
\nonumber \\
\lambda'_{ijk} & \sim & \epsilon^{\tl_i-\tl_0}\lambda^D_{jk} \ , 
\label{general}\\  
\lambda^N_{ij} & \sim & \epsilon^{\tl_i-\tl_0}\lambda^N_{j} \ .
\nonumber
\end{eqnarray} 

Notice that   by the naive  power  counting the   $\epsilon$ factor in
(\ref{general}) would be $\epsilon^{l_i-l_0}$. Therefore, according to
the  values of the charges $l_\alpha$,  we distinguish four different
patterns in the   relations (\ref{general}), as  follows. 
 \vskip .5cm
  (I) $l_i + h_u > l_0 + h_u \ge 0 .$ \\ 
\noindent In this case, all  couplings in (\ref{W1}) are non-vanishing
(unless there are  zeroes in the  corresponding fermion mass matrices)
and, in the combinations       given   by the {\sl     r.h.s.}'s    in
(\ref{lambda}), all terms such that $l_\alpha \le l_i,$ are comparable
in  magnitude. Then  the naive  power  counting  is preserved and  the
relations (\ref{general}) are verified for any  values of the indices,
but the matrices  $\lambda'_{i}$  ($\lambda_{i}$) do not  commute with
the Yukawa $\lambda^E$ (resp.  $\lambda^D)$ matrix. Hence the R-parity
violations    are flavour changing   in  this case. These non-diagonal
lepton number violations  are required to be  extremely small  by FCNC
processes, in particular by the  limits on ${\epsilon}_K $ and $\Delta
m_K$  \footnote{For  a   review   on these limits  see,    {\sl e.g.,}
\cite{Drei}}. However,  in models which  account for realistic fermion
mass hierarchies, it is sufficient to require $l_i-l_0 \ge 3$ in order
to fulfil these   constraints. This suppresses  as  well quark flavour
conserving R-parity violations. 

\vskip .5cm
  (II)  $l_0 + h_u \le 0 < l_i + h_u $  \\ 
\noindent This pattern of charges gives rise to an enhancement of 
flavour conserving\footnote{  Notice  that  the alignment discussed
here means that, for the  mass  eigenstates, the $\lambda'_{ijk}$  matrix
elements in (\ref{general}) are non-vanishing for  $j=k$, and that the
six    non-vanishing       antisymmetric    purely  leptonic couplings
$\lambda_{ijk}$ have $j=k$ or $i=k$. For simplicity, we also refer to
the latter as   flavour conserving  lepton number  violation.}  lepton
number  violation.  Indeed, the  naive power  of  $\epsilon$ would  be
different from that   in    (\ref{general})  since, in   this    case,
$\tl_i-\tl_0 = (l_i-l_0) -  2\tl_0.$ The R-parity violating  couplings
are larger by  a factor $\epsilon^{-  2\tl_0}.$  Furthermore, only the
same  terms with   $\alpha=0$  (or  $\beta  =  0)$  in  (\ref{lambda})
contribute to this enhanced couplings  as well as to the corresponding
Yukawa matrices. For example, 
\begin{equation}
\lambda^D_{jk}  \sim   \lambda^D_{0 j k} \ , \  
\lambda'_{ijk}  \sim  \lambda^D_{0 jk} \epsilon^{\tl_i-\tl_0}
\sim \epsilon^{-2 \tl_0} \epsilon^{l_i+q_j+d_k}  
  \label{align}
\end{equation}
and  similarly for $\lambda^E_{jk}$ and $\lambda_{ijk}$ ($\lambda^N_k$
and $\lambda^N_{ik}$).  Hence,  the matrices of  the  couplings of the
$H_0$  and  all the   $L_i$'s   are approximately  proportional, i.e.,
aligned in the  flavour  space. The  leading R-parity   violations are
predicted to be diagonal in the quark flavours.  This can also be seen
as a  suppression of the  flavour  changing lepton  number violations,
which obey the naive power counting, and a more comfortable fulfilment
of the experimental constraints from FCNC processes. In this case, the
flavour diagonal  couplings $\lambda'_{i}$ and $\lambda_{i}$  might be
larger  by a factor $\epsilon^{-  2\tl_0}$  as compared  to case  (I).
This would  result  in a   factor of $O(10^3)$  for  the corresponding
widths and cross-sections since, as already discussed, $\tl_0 = 1$ for
intermediate values of $\tan{\beta}$. 

\vskip .5cm
(III) $l_i + h_u < l_0 + h_u \le 0 $ \\  
\noindent As explained above, in this case the charges $l_i$ have to be 
sufficiently  negative, so  that  the non-vanishing  couplings  in the
right-hand side of (\ref{lambda})  have $\alpha=0$ or  $\beta=0$.  One
recovers  the  alignment in  flavour space  as  in  (\ref{align}).  The
lepton number   violating trilinear couplings  are  all driven  by the
mixing (\ref{Lhat}) induced by the misalignment  between ${\bf v}$ and
\mubine\ so that  they are fully aligned to  the $H_d$  couplings. The
power  of $\epsilon$ in (\ref{general}) is  the opposite  of the naive
counting   one and the R-parity   violations can be  suppressed by the
choice of the charge differences, $l_i - l_0,$ and are not constrained
by $K\bar{K}$ mixing. 

\vskip .5cm
(IV) $l_i + h_u < 0  \le l_0 + h_u  $ \\  
\noindent If the charges $l_i$ are not negative enough to imply vanishing 
couplings by analyticity,  the lepton number  violating couplings obey
the naive power    counting while the  Higgs couplings   would  get an
enhancement factor of $\epsilon^{-2\tl_0}$.  There is no alignment and
the R-parity violating couplings  still supersede the Higgs  couplings
by a factor $\epsilon^{\tl_i-\tl_0}$ in spite of their enhancement. In
order to satisfy the phenomenological constraints, the $l_i$'s have to
be more negative, and the pattern of lepton number violating couplings
will be    similar  to case   (III)  above,  with  alignment given  in
(\ref{align}) and the failure of naive power counting. 

Therefore, in models with abelian charge  assignments that satisfy the
experimental requirements, the lepton  number violating couplings obey
the naive  power counting  in case (I),  but not   in the other  three
cases, where they are aligned to the fermion  mass matrices in flavour
space.  Moreover, in case (II) this property  is due to an enhancement
of flavour diagonal R-parity violation! 

An important consequence of lepton number  violation is the generation
of neutrino masses. At  the tree-level, one neutrino   gets a mass  as
given by  (\ref{eq:mLHu}) and (\ref{m0}).  If the gauge singlets $N_i$
are  not introduced in  the theory, the  other two neutrino states get
their masses and mixings at the one loop level. This has been recently
discussed \cite{BoGNN} in detail for case (I), but the other cases are
quite   similar  (with   the  assumption  made  in    case (III)). For
completeness, we just present the  general expression for the neutrino
mass matrix,  which takes into account  the loop contributions as well
as the seesaw masses.  It can be written as follows: 
\begin{equation}
(m_{\nu})_{ij} = \epsilon^{\tl_i+\tl_j-2\tl_0}(m_0 \delta_{i3}
\delta_{j3} + m_{{\rm loop}} + m_{{\rm seesaw}}).
\label{mnu}
\end{equation} 
where $m_{{\rm loop}}$ is a scale defined by the loop contributions 
\cite{HS,loop}, which are
dominated by the $b\tilde{b}$ one, so that \cite{BoGNN},
\begin{equation}
m_{{\rm loop}} \sim {{ 1\ {\rm keV}}\over{\cos^2{\beta}}} \ \left(
{{500\; {\rm GeV}}\over{\tilde{m}}}\right),
\label{loop}
\end{equation} 
where $\tilde{m}$ is the squark mass. The seesaw contribution
corresponds to the scale,
\begin{equation}
m_{{\rm seesaw}} \sim ( 1\ {\rm eV})\epsilon^{2(l_0+h_u)} \ 
\left( {{10^{13}\; {\rm GeV}}\over{M_R}}\right),
\label{seesaw}
\end{equation} 
where  we have introduced  the large scale  $M_R$  such that the $N_i$
mass matrix  in the superpotential (\ref{W1})  is of the  form $M_{ij}
\sim \epsilon^{n_i+n_j} M_R.$ For $M_R >  10^{10} {\rm GeV}$, the loop
contribution  dominates over  the seesaw  mechanism.  For  the sake of
comparison, we approximate (\ref{m0}) by 
\begin{equation}
m_0 \sim { (100\ {\rm GeV})}\cos^2{\beta} \ \left( {{500\; {\rm GeV}}
\over{\tilde{m}}}\right),
\label{maprox}
\end{equation} 
and we recall that  $\cos{\beta}  \sim \epsilon^{\tl_0}.$ In order  to
satisfy the  cosmological limits  on the  $\nu_{\tau}$ mass,  we  must
require   $\tl_3 \ge  7.$  We  refer to   \cite{BoGNN}  for a detailed
phenomenological discussion of these  predictions, which extend to the
three cases above as already remarked. 

It is worth noticing  that  the powers of  $\epsilon$ in  the neutrino
mass  matrix (\ref{mnu}) are  the  same  that  appear in the  relation
between lepton number violating couplings and  the Yukawa couplings in
(\ref{general}), providing relations  for   the magnitudes of    these
physical quantities. For example
\begin{equation}
m_{\nu_3} \sim m_0 ({{\lambda'}_{3jk} \over {\lambda^D_{jk}}})^2 \ . 
\end{equation}

\vskip .5cm
We now come back to the problem of baryon number violation in this class
of models. We already noticed that there is a qualitative difference
between the couplings $\lambda^{''}$ and $\lambda, \lambda'$. The latter,
even if for some symmetry reason they are absent from the
superpotential, can be generated for leptons and down quarks 
through the Higgs-lepton mixing that we
discussed previously. On the contrary, if a symmetry reason forbids the
former ($\lambda^{''}$) couplings in the superpotential and allow them
in the K\"ahler potential, they will only appear after supersymmetry
breaking through the Giudice-Masiero mechanism \cite{GM}. More precisely, if 
$u_i+d_j+d_k <0$ for all $i,j,k=1,2,3$, then at high energy we have
terms in the K\"ahler potential of the type ${1 \over M_P} 
{\bar \epsilon}^{-(u_i+d_j+d_k)} U^i D^j D^k$. Then 
we get contributions in the low-energy effective superpotential 
\begin{equation}
W \sim {{\tilde m} \over M_P} \epsilon^{|u_i+d_j+d_k|} U^i D^j D^k \ ,
\label{baryon} \
\end{equation}
where $M_P$ is the Planck mass. We therefore find suppressed couplings,
of order ${10}^{-17}$ to ${10}^{-19}$ depending on the specific model, for
moderate negative quark charges. Combining these values with the
corresponding ones for $\lambda'$ as  discussed previously, we find that the
proton decay can be suppressed down to acceptable values which 
could be tested in the forthcoming years. Of course, not all the couplings
(\ref{baryon}) are equally dangerous and this discussion can be refined in a
specific model.

Because of the usual quark Yukawa couplings, this mechanism generally asks for
large positive  $q_i$ charges and is constrained by the
$U(1)$ anomaly cancellation conditions. We have searched for explicit
solutions in models based on a family $U(1)$ symmetry \cite{LNS},
\cite{IR}, \cite{BR} with anomaly cancellation \`a la Green-Schwarz \cite{GS}.
By imposing anomaly conditions and using explicit models \cite{BR} , we
found that the mechanism can be implemented in cases (III) and (IV)
with the standard particle content. We give as an illustration one model   
with the following charge assignements:
\begin{eqnarray} 
q_1=6 \ , q_2=5 \ , q_3=3 \ , u_1=7 \ , u_2=4 \ , u_3=2 \ , \nonumber \\
d_1=-7 \ , d_2=-8 \ , d_3=-8 \ , l_1=-8 \ , l_2=-8 \ , l_3= -8 \ .
\end{eqnarray}
In this example, the relevant couplings mediating proton decay satisfy
$\lambda' \lambda'' \le 10^{-16} \epsilon^{16} \sim 10^{-26}$ for $\epsilon
\simeq 0.22$, which shows the high degree of 
suppression which can be obtained if such couplings are obtained through the
Giudice-Masiero mechanism. We emphasize that if we relax
the anomaly cancellation conditions (by allowing the presence of exotic
particles in the spectrum) models can be proposed with more moderate 
values of the charges
and efficient suppression of proton decay.

\end{document}